# Complex refractive index variation in proton-damaged diamond

S. Lagomarsino,[1,*] P. Olivero,[2,3,4] S. Calusi,[5] D. Gatto Monticone,[2,3] L. Giuntini,[5] M. Massi,[5] S. Sciortino,[1] A. Sytchkova,[6] A. Sordini,[7] and M. Vannoni,[7]

1 *Energetics Department and INFN Sezione di Firenze, University of Firenze, via Sansone 1, 50019 Sesto Fiorentino, Firenze, Italy*
2 *Experimental Physics Department and "Nanostructured Interfaces and Surfaces" Centre of Excellence, University of Torino, via P. Giuria 1, 10125 Torino, Italy*
3 *INFN Sezione di Torino, via P. Giuria 1, 10125 Torino, Italy*
4 *Consorzio Nazionale Interuniversitario per le Scienze Fisiche della Materia (CNISM) Italy*
5 *Physics Department and INFN Sezione di Firenze, University of Firenze (Italy), via Sansone 1, 50019 Sesto Fiorentino, Firenze, Italy*
6 *ENEA, Optical Coatings Laboratory, via Anguillarese 301, 00123 Roma, Italy*
7 *CNR, Istituto Nazionale di Ottica (INO), Largo E. Fermi 6, 50125 Arcetri, Firenze Italy*
*[lagomarsino@fi.infn.it](lagomarsino@fi.infn.it)*

**Abstract:** An accurate control of the optical properties of single crystal diamond during microfabrication processes such as ion implantation plays a crucial role in the engineering of integrated photonic devices. In this work we present a systematic study of the variation of both real and imaginary parts of the refractive index of single crystal diamond, when damaged with 2 and 3 MeV protons at low-medium fluences (range: $10^{15}$ - $10^{17}$ cm$^{-2}$). After implanting in 125×125 μm$^2$ areas with a scanning ion microbeam, the variation of optical pathlength of the implanted regions was measured with laser interferometric microscopy, while their optical transmission was studied using a spectrometric set-up with micrometric spatial resolution. On the basis of a model taking into account the strongly non-uniform damage profile in the bulk sample, the variation of the complex refractive index as a function of damage density was evaluated.

### References and links

1. P. Kok and B. W. Lovett, "Qubits in the pink," Nature **444**, 49-49 (2006)
2. I. Aharonovich, C. Zhou, A. Stacey, J. Orwa, S. Castelletto, D. Simpson, A. D. Greentree, F. Treussart, J.-F. Roch, and S. Prawer, "Enhanced single-photon emission in the near infrared from a diamond color center," Phys. Rev. B **79**, 235316 (2009)
3. C. Wang, C. Kurtsiefer, H. Weinfurter, and B. Burchard, "Single photon emission from SiV centres in diamond produced by ion implantation," J. Physics B **39**, 37-41 (2006)
4. I. Aharonovich, S. Castelletto, B. C. Johnson, J. C. McCallum, D. A. Simpson, A. D. Greentree, and S. Prawer, "Chromium single-photon emitters in diamond fabricated by ion implantation," Phys. Rev. B **81**, 121201 (2010)
5. A. Beveratos, R. Brouri, T. Gacoin, A. Villing, J-P. Poizat, and P. Grangier, "Single photon quantum cryptography," Phys. Rev. Lett. **89**, 187901 (2002)
6. S. Prawer, and A. D. Greentree, "Diamond for quantum computing," Science **320**, 1601-1602 (2008)

7. M. P. Hiscocks, C.J. Caalund, , F. Ladouceur, S. T. Huntington, B. C. Gibson, S. Trpkovscki, D. Simpson, E. Ampem-Lassen, S. Prawer, and J. E. Butler, "Reactive ion etching of waveguide structures in diamond," Diamond Relat. Mater. **17**, 1831-1834 (2008)
8. M. P. Hiscocks, K. Ganesan, B. C. Gibson, S. T. Huntington, F. Ladouceur, and S. Prawer, "Diamond waveguides fabricated by reactive ion etching," Opt. Expr. **16** (24), 19512-19519 (2008)
9. C. F. Wang, R. Hanson, D. D. Awschalom, E. L. Hu, T. Feygelson, J. Yang, and J. E. Butler. "Fabrication and characterization of two-dimensional photonic crystal microcavities in nanocrystalline diamond," Appl. Phys. Lett. **91**, 201112 (2007)
10. B. A. Fairchild, P. Olivero, S. Rubanov, A. D. Greentree, F. Waldermann, R. A. Taylor, I. Walmsley, J. M. Smith, S. Huntington, B. C. Gibson, D. N. Jamieson, and S. Prawer, "Fabrication of Ultrathin Single-Crystal Diamond Membranes," Adv. Mater. **20**, 4793-4798 (2008)
11. C. F. Wang, Y-S. Choi, J. C. Lee, E. L. Hu, J. Yang, and J. E. Butler, "Observation of whispering gallery modes in nanocrystalline diamond microdisks," Appl. Phys. Lett. **90**, 081110 (2007)
12. S. Lagomarsino, P. Olivero, F. Bosia, M. Vannoni, S. Calusi, L. Giuntini, and M. Massi, "Evidence of Light Guiding in Ion-Implanted Diamond," Phys. Rev. Lett. **105**, 233903 (2010)
13. S. Snjezana Tomljenovic-Hanic, A. D. Greentree, C. Martijn de Sterke, and S. Prawer, "Flexible design of ultrahigh-Q microcavities in diamond-based photonic crystal slabs," Opt. Express **17**(8), 6465-6475 (2009)
14. E. Gu, H. W. Choi, C. Liu, C. Griffin, J. M. Girkin, I. M. Watson, M. D. Dawson, G. McConnell, and A. M. Gurney, "Reflection/transmission confocal microscopy characterization of single-crystal diamond microlens arrays," Appl. Phys. Lett. **84**(15), 2754-2756 (2004)
15. Y. Fu and N. K. A. Bryan, "Investigation of diffractive optical element fabricated on diamond film by use of focused ion beam direct milling" Opt. Eng. **42**(8), 2214 (2003)
16. S. Gloor, V. Romano, W. Lüthy, H. P. Weber, V. V. Kononenko, S. M. Pimenov, V. I. Konov, and A. V. Khomich, "Antireflection structures written by excimer laser on CVD diamond," Appl. Phys. A **70**, 547-550 (2000)
17. H. Björkman, P. Rangsten, and K. Hjort, "Diamond microstructures for optical micro electromechanical systems," Sens. Act. **78**, 41-47 (1999)
18. R. L. Hines, "Radiation Damage of Diamond by 20-keV Carbon Ions," Phys. Rev. **138**(6A), A1747 (1965)
19. M. G. Jubber, M. Liehr, J. L. McGrath, J. I. B. Wilson, I. C. Drummond, P. John, D. K. Milne, R. W. McCullough, J. Geddes, D. P. Higgins, and M. Schlapp, "Atom beam treatment of diamond films," Diamond Relat. Mater. **4**, 445-450 (1995)
20. A. Battiato, F. Bosia, S. Ferrari, P. Olivero, A. Sytchkova, and E. Vittone, "Spectroscopic measurement of the refractive index of ion-implanted diamond" Opt. Lett. **37**, 671-673 (2012)
21. P. Olivero, S. Calusi, L. Giuntini, S. Lagomarsino, A. Lo Giudice, M. Massi, S. Sciortino, and M. Vannoni, "Controlled variation of the refractive index in ion-damaged diamond," Diamond Relat. Mater. **19**, 428-431 (2010)
22. K. L. Bathia, S. Fabian, S. Kalbitzer, C. Klatt, W. Krätschmer, R. Stoll, and J. F. P. Sellschop, "Optical effects in carbon-ion irradiated diamond," Thin Solid Films **324**, 11-18 (1998)
23. A. V. Khomich, V. I. Kovalev, E. V. Zavedeev, R. A. Khmelnitskiy, and A. A. Gippius, "Spectroscopic ellipsometry study of buried graphitized layers in the ion-implanted diamond," Vacuum **78**, 583-587 (2005)
24. A. A. Bettiol, S. V. Rao, E. J. Teo, J. A. van Kan, and F. Watt, "Fabrication of buried channel waveguides in photosensitive glass using proton beam writing," Appl. Phys. Lett. **88**, 171106 1-3 (2006)
25. F. Chen, L. Wang, Y. Jiang, X.-L. Wang, K.-M. Wang, G. Fu, Q.-M. Lu, C. E. Rüter, and D. Kip, "Optical channel waveguides in Nd: YVO4 crystal produced by O+ ion implantation," Appl. Phys. Lett. **88**, 071123 1-3 (2006)
26. L. Giuntini, M. Massi, and S. Calusi, "The external scanning proton microprobe of Firenze: a comprehensive description," Nucl. Instrum. Meth. A **576**, 266-273 (2007)
27. K. Creath, "Phase-shifting interferometry techniques", in *Progress in Optics* (Elsevier, 1988)
28. A. K. Sytchkova, J. Bulir, and A. M. Piegari, "Transmittance measurements on variable coatings with enhanced spatial resolution" Chinese Opt. Lett. **8**, 103-104 (2010)
29. J. F. Ziegler, J. P. Biersack, and U. Littmark, *The Stopping and Range of Ions in Solids* (Pergamon, 1985)
30. E. W. Maby, C. W. Magee, and J. H. Morewood, "Volume expansion of ion-implanted diamond," Appl. Phys. Lett. **39**, 157-158 (1981)
31. J. F. Prins, T. E. Derry, and P. F. Sellschop, "Volume expansion of diamond during ion implantation," Phys. Rev. B **34**(12), 8870-8874 (1986)
32. M. Massi, L. Giuntini, M. Chiari, N. Gelli, and P.A. Mandò, "The external beam microprobe facility in Florence: set-up and performance," Nucl. Instr. and Meth. in Phys. Res. B **190**, 276-282 (2002)
33. L. Giuntini, M. Massi, and S. Calusi, "The external scanning proton microprobe of Firenze: a comprehensive description," Nucl. Instr. and Meth. in Phys. Res. A **576** 266-273 (2007)
34. P.A. Mandò, "Advantages and limitations of external beams in applications to arts & archeology, geology and environmental problems," Nucl. Instr. and Meth. in Phys. Res. B **85**, 815-823 (1994)
35. T. Calligaro, J.-C. Dran, E. Ioannidou, B. Moignard, L. Pichon, J. Salomon, and "Development of an external beam nuclear microprobe on the Aglae facility of the Louvre museum," Nucl. Instr. and Meth. in Phys. Res. B **161**, 328-233 (2000)

36. P.A. Mandò, "Measurement of low currents in an external beam set-up," Nucl. Instr. and Meth. in Phys. Res. B **85**, 815 (1994)
37. M. Chiari, A. Migliori and P.A. Mandò, "Investigation of beam-induced damage to ancient ceramics in external-PIXE measurements," Nucl. Instr. and Meth. in Phys. Res. B **188**, 162-165 (2002)
38. L. Giuntini, "A review of external microbeams for ion beam analyses", Anal. Bioanal. Chem. **401** (3) 785-793 (2011).
39. M. Vannoni, G. Molesini, S. Sciortino, S. Lagomarsino, P. Olivero and E. Vittone, "Interferometric characterization of mono-and polycrystalline CVD diamond," Proc. SPIE **7389**, 738931-1_6 (2009).
40. J. H. Bruning, "Fringe Scanning Interferometers,"in *Optical Shop Testing*, (Wiley, 1978).
41. J.F. Ziegler, M.D. Ziegler, and J.P. Biersack, "SRIM – The stopping and range of ions in matter (2010)," Nucl. Instr. and Meth. in Phys. Res. B **268**, 1818-1823 (2010)
42. W. Wu and S. Fahy, "Molecular-dynamics study of single-atom radiation damage in diamond," Phys. Rev. B **49**, 3030-3035 (1994)
43. C. Uzan-Saguy, C. Cytermann, R. Brener, V. Richter, M. Shaanan, and R. Kalish, "Damage threshold for ion-beam induced graphitization of diamond," Appl. Phys. Lett. **67**, 1194-1196 (1995)
44. P. Olivero, S. Rubanov, P. Reichart, B. C. Gibson, S. T. Huntington, J. R. Rabeau, A. D. Greentree, J. Salzman, D. Moore, D. N. Jamieson, and S. Prawer, "Characterization of three-dimensional microstructures in single-crystal diamond," Diamond Relat. Mater. **15**, 1614-1621 (2006);]
45. D. P. Hickey, K. S. Jones, and R.G. Elliman, "Amorphization and graphitization of single-crystal diamond - a transmission electron microscopy study," Diamond Relat. Mater. **18**, 1353-1359 (2009)
46. R. L. Hines, and R. Arndt, "Radiation effects of bombardment of quartz and vitreous silica by 7.5-kev to 59-kev positive ions," Phys. Rev. **119**, 623-633 (1960)
47. L. Babsail, N. Hamelin, and P. D. Townsend, "Helium-ion implanted waveguides in zircon," Nucl. Instr. and Meth. in Phys. Res. B **59/60**, 1219-1222 (1991)
48. D. T. Y. Wei, W. W. Lee, and L. R. Bloom, "Large refractive index change induced by ion implantation in lithium niobate," Appl. Phys. Lett. **25**, 329-331 (1974)

---

## 1. Introduction

In the search of a reliable platform for a scalable fabrication technology of quantum devices, diamond has been attracting growing interest due to a number of remarkable properties.
High brightness of impurity-related (N, Si, Ni, Cr) color centers gives indication of high dipole moment and strong coupling with electromagnetic field, allowing effective applications in single photon sources [1-4]. The null magnetic moment of the C12 nucleus allows the coherence time of the NV— sublevels of the triplet ground state to be very long, making them candidates for quantum bit storage even at room temperature [5,6].
Since the formation of active optical centers in diamond is inherently related to the creation of crystal structure defects, a suitable control on the variation of the refractive index as a function of structural damage/disorder is highly required in advanced photonics applications. With the aim of exploiting the above-mentioned attracting properties, several diamond micro-fabrication methods are under study [7-12], promising to offer a viable path towards the integration of monolithic photonic devices while exploiting the broad-band transparency and high refractive index of this material. Such methods are often based on ion-beam microfabrication strategies [7,9,11,12]: possible variations of the refractive index due to structural damage during the device fabrication process must be accurately predicted to properly design the devices of interest . Moreover, with the aim of fabricating photonic devices in bulk diamond, the low-contrast refractive index modulation induced by ion implantation, instead of merely being a side effect, could play an active role in a more effective device design [12,13].
Finally, a suitable control of the optical properties of damaged diamond is demanded also in a broad range of more conventional micro-optics applications, e.g. high-power laser windows and lenses, optical MEMS, optical data storage [14-17].
The effect of ion-beam induced structural damage on the refractive index in diamond has been observed since the '60 [18] and qualitatively reported in the literature[19]. In spite of this, remarkably only few works were dedicated to its systematic investigation [20-22]. One example is reported in ref. [22], where carbon ions of different energies (50 keV – 1.5 MeV) were subsequently implanted in the same area, in order to produce a homogeneous damage profile over a depth of 1 µm; the refractive index was then measured on the as-implanted samples as a function of the implantation fluence, by means of reflectometric methods. In ref. [23] an ellipsometric study is reported in which the refractive index is measured from heavily damaged buried graphitic layers produced in diamond with 350 keV He+ ion irradiation.

Monoenergetic implantations with MeV light ions, such as hydrogen or helium, create damage profiles significantly different from those reported in the previous examples, because they induce the formation of modified regions lying deeper under the diamond surface, whose characterization with reflectometric methods is much more difficult. Nonetheless, the employment of MeV light ions can be an extremely versatile tool to locally modify the optical properties of materials with micrometric spatial resolution both in the lateral and depth directions, thanks respectively to the above-mentioned peculiar damage profile and to the possibility of focusing MeV ion beams to the micrometer scale with electromagnetic lenses. The strong potential of MeV ion microbeam implantation for the direct writing of optical structures has already been demonstrated in other materials of technological interest [24,25], and recently proved also in diamond [12].

In our study, IIa monocrystalline diamonds grown by Chemical Vapor Deposition (CVD) were implanted with a scanning microbeam of 2 and 3 MeV protons [26], at fluences in the $10^{15}$-$10^{17}$ cm$^{-2}$ range. The damaged regions lie respectively 24 and 48 µm below the diamond surface and extend for few (i.e. 2-6) micrometers.

In order to measure the damage-induced variations of refractive index and absorption coefficient, an interferometric transmission microscopy technique [27] and a space-resolved transmission spectroscopic setup [28] were employed. The probe light wavelength was 632.8 nm, conveniently close to the zero-phonon-line emission of the NV– center (637 nm), arguably the most widely investigated color center in diamond for applications in quantum optics. In order to estimate the variation of the real and imaginary parts of the refractive index as a function of damage density, the direct measurements of the optical path difference (OPD) and the difference in absorption length (absorption length difference, ALD) between implanted and unimplanted regions were interpreted with a phenomenological model developed by the authors, based on the damage depth profile obtained with Monte Carlo SRIM simulations [29], and then compared with a full multilayer propagation model.

We have also tested the possible dependence of the variation of the optical properties of the material from other implantation parameters, such as fluence delivery rate, i.e. ion beam current, to exclude self-annealing effects, and the incidence angle, in order to verify the possible influence of channeling effects.

The samples under investigation are described in section 2.1, while section 2.2 is dedicated to the description of the ion implantation process. In section 2.3 the measurement methods for the determination of the OPD and ALD are outlined, together with the measurement method of the surface deformation (swelling) due to the expansion of the damaged regions [30,31]. In section 3 the data analysis is presented, along with the description of the interpretation model, and the final results are presented in term of the dependence of the complex refractive index on the damage level, i.e. the density of vacancies produced by ion irradiation.

## 2. Experimental results

### 2.1 Samples

This study was carried out on five 3.0×3.0×0.5 mm$^3$ single-crystal diamonds grown with Chemical Vapour Deposition (CVD) technique by ElementSix (http://www.e6cvd.com). The samples consist of a single {100} growth sector and are classified as type IIa, with concentrations of nitrogen and boron impurities below 0.1 ppm. The crystals were cut along the <100> axes and the two opposite faces of the samples were optically polished.

### *2.2 Ion implantation*

The diamond samples were implanted at the external scanning microbeam facility [32,33] (Fig. 1) of the 3 MV Tandetron accelerator of the INFN LABEC Laboratory in Florence.
The diamond to be implanted was kept out of vacuum, thus allowing its easy handling, positioning and monitoring [34]. Before hitting the target, the beam passes through a thin silicon nitride ($Si_3N_4$) membrane, 100 nm thick and 1×1 mm2 wide [35] (inset of Fig.1), sealing the final part of the vacuum line, and 2 mm of unenclosed helium atmosphere. The extreme thinness of $Si_3N_4$ window and the short external path in helium allow to minimize beam widening and energy straggling; as a result, 10-20 µm spot size on sample is obtained, with ~10 keV of energy straggling for MeV protons.
A magnetic beam-scanning system was used to control the position of the beam impact point on the sample within a ~1×1 $mm^2$ area, corresponding to the exit window aperture. A multi-axis linear motorized stage of a 25 mm range allows high resolution translation of the sample on the plane normal to beam axis, with position reproducibility better than 1 µm.

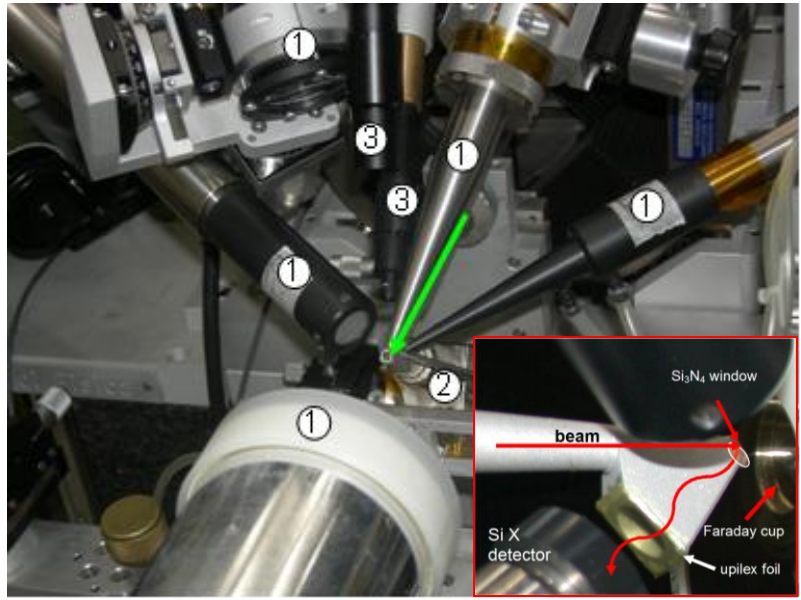

**Fig. 1.** Exit nozzle of the ion microbeam line: (1) detectors for Ion Beam Analysis (not used in this application) (2) X-ray detector for beam charge measurement, (3) vista camera and micro camera; the arrow indicates the ion beam direction. Inset: Details of the system for beam charge measurement.

For this study, proton beams were focused on the polished side of the samples to a spot of ~10 µm (3 MeV) and ~20 µm (2 MeV). Different zones of the samples were implanted at fluences ranging from ~$10^{15}$ $cm^{-2}$ to ~$10^{17}$ $cm^{-2}$. For each implantation, the ion beam was magnetically scanned exploiting the same raster frame of ~125×125 µ$m^2$, much wider than the beam spot dimensions, in order to deliver a homogeneous fluence over a vast central area of each irradiated zone.
During the implantations, fluences were determined by measuring the implanted charge (i) and setting the size of the irradiated area (ii), as described hereafter.

(i) *Implanted charge*: we used the beam charge measuring system installed at the LABEC microbeam, which exploits the yield of Si X-rays produced by the beam in the exit window [34]. The total charge implanted into the sample can be expressed as $Q_i = K \cdot A_{X-Si}$, being $K$ a proportionality factor and $A_{X-Si}$ the number of Si X-rays counted by a dedicated detector, as reported in detail in ref. [26]. The calibration factor $K$ was determined, for two samples, by measuring the ratio of the integrated charge ($Q_I$), collected with a Faraday cup [36] surrounding the exit nozzle (Fig. 1), to the Si X-ray yield ($A_{X-Si}$). For the other samples, the factor $K$ was evaluated by comparing the time-integrated X-ray yield with the back-scattered proton fluence from a gold target. In the whole explored range of beam currents (0.2 - 1.5 nA), K remained constant within ~1% of its value. As a result, the overall precision on the implanted charge determination is ~1%, being the statistical error related to the Si X-rays counting typically well below 1%. Possible systematic errors in the charge determination, affecting all the experimental points with a common scale factor, are ~10% [37].

(ii) *On-line setting of irradiated area*: In order to implant the ions in areas of controlled dimensions, we calibrated the magnetic displacement of the beam on the sample surface by exploiting a standard TEM Cu grid. The uncertainty on the scanned area, which is basically due to the calibration procedure, is ~5%. After ion implantation, the size of the irradiated area was measured on the OPD maps as described in Section 3, thus improving the precision on the area determination up to ~2%.

The visual aspect of the sample after the process is shown in Fig. 2. It is apparent the darkening due to ion damage of the implanted areas.

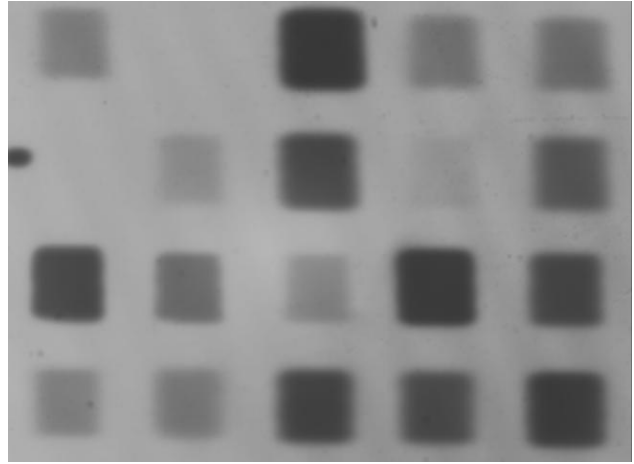

**Fig. 2**. Transmission optical image of several 125×125 μm² implanted areas. Progressive darkening of the implantation regions with increasing total fluence, along with fading of induced ion-luminescence [38], allows a qualitative control of the implantation progress.

*2.3 Optical characterisation*

In order to evaluate the variation of the refractive index due to ion-induced damage, the phase shift of a laser beam crossing the damaged diamond layer was determined using a commercial laser interferometric microscope (Maxim 3D, Zygo Corporation, Middlefield, CT, USA) with a 20× micro-Fizeau objective, operating in the He-Ne 632.8 laser line, with horizontal and vertical resolutions of 1.68 μm and 0.63 nm, respectively, and with a field view of 349×317 μm. The instrumental setup is schematically summarized in Fig.3 [39].

A He-Ne laser beam is properly expanded to invest the full area of the sample; the micro-Fizeau objective contains a beam-splitter that reflects part of the light ("reference beam"), while the remaining part crosses the sample and is reflected from a high-quality external mirror ("test beam"). The diamond is slightly tilted to avoid undesired internal reflections between the two opposite surfaces of the sample. The interference pattern of the reference and test beam is recorded by a CCD camera.

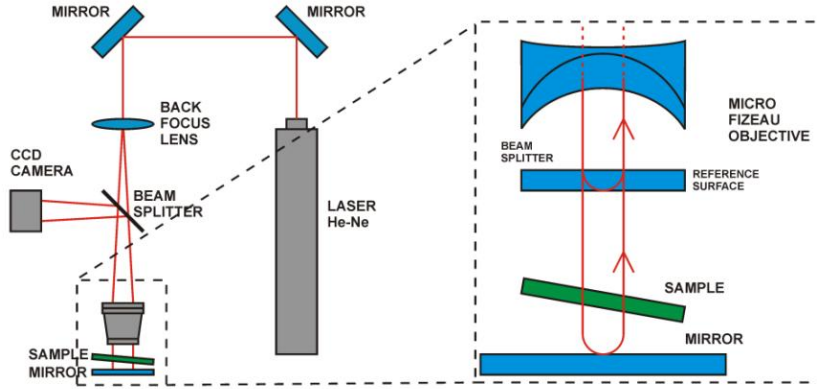

**Fig. 3**. Schematics of the experimental principle of the Zygo interferometric microscope.

Using the phase shift method [40] it is possible to reconstruct the relative phase $\Delta\phi$ of the test beam at each pixel: the contributions of the beam splitter and the high-quality mirror is accounted for and removed. The phase difference $\Delta\phi$ reflects the optical path difference (OPD) between the light crossing the whole implanted region and the un-damaged one (see Fig. 4(a)).

There is also a smaller contribution to the phase difference (about 15% of the OPD, see Fig. 4(b)) due to the expansion ("swelling") of the highly damaged layer è [30,31], which was measured with a white light interferometric profilometer (Zygo NewView). The contribution of swelling to the OPD signal has been calculated and properly deconvoluted from the part responsible for the variation of the refractive index only, as reported in Section 3.

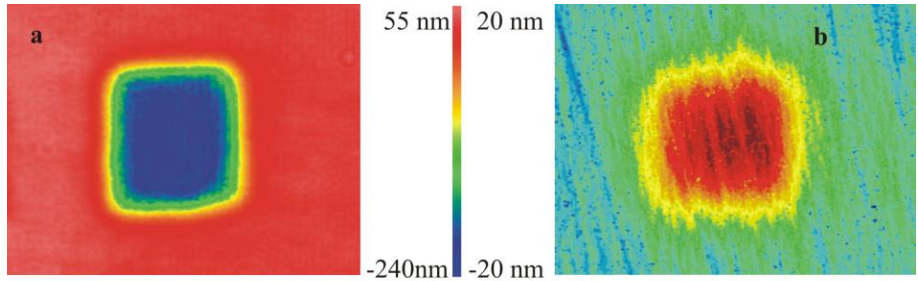

**Fig. 4**. Schematics a) Map of the OPD profile of an area implanted with 3 MeV protons at a fluency of $7.5 \cdot 10^{16}$ cm$^{-2}$; b) map of the swelling profile of the same implantation. The two profiles have opposite signs because swelling gives a shorter optical path in reflection, while damage produces a longer optical path in transmission measurements.

The optical absorption of the irradiated zones was estimated from their transmittance values obtained with a custom set up for measurements with high spatial resolution [28]. The light of a Xe-source is guided by a 5μm fiber optic wire, forming a spot zone on the sample surface of ~50 μm, which determines the spatial resolution of the system selected for this study. Subsequently, the transmitted light was focused on a second optical fiber, connected to an Ocean Optics spectrometer SQ2000 having a spectral resolution of 0.8 nm and spectral range

400-1200 nm. The finite spot size of the incident beam may widen any narrow spectral features if the transmittance varies very rapidly across the sample surface. However, from the OPD measurements the implanted region results to be uniform within an area much wider than the beam spot, so that a spectral resolution of at least 1 nm is guaranteed. The spectra were acquired at the position of minimum transmittance, within the area of implantation.

In the present work the absorption values were estimated at the same wavelength employed in the OPD measurement ($\lambda_{laser}$ = 632.8 nm); a full spectral analysis of the optical absorption data will be the subject of a forthcoming publication.

## 3. Data analysis

### 3.1 Dependence on fluence of the OPD and the ALD

The optical path difference between the center of the implanted area and the surrounding unimplanted region is estimated by the difference between the *OPD* mean value in a central square region and in a frame region located respectively well inside and outside the irradiated area. The uncertainty of the *OPD* measure, evaluated by the fluctuations of the phase inside and outside each region, is between 3 and 10 nm, which is predictably of the same order of magnitude of the roughness of the diamond surface (~2 nm) multiplied by the refractive index difference between diamond and air at the probed wavelength (~1.41).

The absorption length difference was evaluated, for each implantation, by the ratio between the transmittance $T_0$ of un-implanted substrate, i.e. of the pristine diamond, and the value $T$ measured at a chosen damaged area:

$$ALD = \frac{\lambda_{He-Ne}}{4\pi} \ln\left(\frac{T_0}{T}\right). \tag{1}$$

Both the *OPD* and the *ALD* measurements are affected by swelling, the expansion of the implanted material determining both a further phase shift of the probe laser beam and an additional absorption contribution. At the lowest order in the displacement of each layer in diamond and in the relative variation of refractive index, the values of OPD and ALD to the net of the swelling effect are obtained by the measured ones ($OPD_m$, $ALD_m$), by the simple equations:

$$OPD = OPD_m - (n_0 - 1)h, \quad ALD = OPD_m - k_0 \cdot h, \tag{2}$$

were $n_0$ and $k_0$ are the refractive index and the extinction coefficient of undamaged pristine diamond and h is the swelling height.

While the extinction coefficient of pristine diamond at 632.8 nm can be assumed to be null and thus the contribution of the term $k_0 \cdot h$ can be neglected, the product $(n_0 - 1)h$ amounts to about 15% of the measured *OPD*, and has been properly subtracted.

The fluence in the central region of each implantation has been calculated simply as the ratio of the deposited charge $Q$ to the area $A_\Omega$ of the raster scanning area. This approximation is justified if the scanning is uniform and the dimensions $L$ of the scanned area is much wider than the beam cross-section l (in our case, $L = 125$ μm >> $l = 10$-$20$ μm).

The charge Q is evaluated by means of the procedure outlined in section IIb with an accuracy of the order of 1%, while $A_\Omega$ is measured directly on the *OPD* maps by evaluating the number of pixels whose *OPD* is above the average value between the *OPD* inside and outside the implanted area. We verified in this way the repeatability of the area setting to be significantly better than the calibration uncertainty obtained with the TEM Cu grid, allowing to keep the overall fluence uncertainty as low as ~3%.

The optical depth and absorption length variation, extracted by the experimental data as illustrated before, reveals a clear correlation with the implantation fluence and ion energies, as shown in Figs. 5 and 6.

### 3.2 Simulation of the ion damage

In order to extract the refractive index variations with the ion-induced damage from the fluence dependence of the *OPD* and *ALD* we need a model which, for any given ion-energy and fluence, gives a physical quantity expressing the entity of damage at a given depth into the diamond. We assumed this quantity to be the induced vacancy density $v(z)$, admitting this parameter to bring all the essential information about the damage processes of a specific ion species and energy. We evaluated $v(z)$ numerically using Monte Carlo SRIM simulations [41], averaged over ensembles of 50,000 ions, by setting the atomic displacement energy o 50 eV [42] and adopting the quick calculation mode.

For an ion of energy E, the simulation provides the number of vacancies per unit length at a depth z as $p^E(z)$ (see Fig. 7). Then, we calculated the induced vacancy density at a given fluence and energy to be $v(z) = \phi \cdot p^E(z)$, supposing that non-linear processes such as self-annealing, ballistic annealing and defect interaction could be neglected. Infact, it has been shown that at damage densities that do not exceed the graphitization threshold (i.e. $1 \cdot 10^{22}$ vacancies cm$^{-3}$ for shallow implantations [43] and $6\text{-}9 \cdot 10^{22}$ vacancies cm$^{-3}$ for deep implantations [44,45] such an hypothesis is valid and provides an adequate description of the ion-induced damage process in diamond in many respects.

### 3.3 Phenomenological model

Let us assume the complex refractive index $\hat{n} = n + ik$ to be directly determined by the vacancy density $v(z)$; here we assume a linear behaviour of kind:

$$\hat{n}(z) = \hat{n}_0 + \hat{c} \cdot v(z) \tag{3}$$

Let us suppose that the complex optical path difference $COPD = OPD + iALD$ between the irradiated and unimplanted areas is exclusively determined by the refractive index as follows:

$$COPD = \int_0^{+\infty} [\hat{n}(z) - \hat{n}_0] \, dz, \tag{4}$$

thus neglecting internal reflections between adjacent differently damaged layers in diamond, and in general considering the processes of refraction and absorption of the probe laser beam as independent from each other. Indeed, it is reasonable to exclude discontinuities along the depth of the irradiated media which might provoke interference phenomena at interfaces; however we may not exclude *a priori* a relatively sharp gradient of the modified refractive index that might induce interference-like effect, and, if the absolute *k* value becomes high enough, might result in *n* and *k* reciprocal dependence. The validity of our supposition has been validated *a posteriori* by means of a full multi-layer optical calculation, as described in the following section.

From Eqs. 3 and 4 the complex optical path differences $COPD^E(\phi)$ at fluence $\phi$ and energy *E* are given by:

$$COPD^E(\phi) = c \cdot I^E \cdot \phi \quad \text{with} \quad I^E = \int_0^{+\infty} p^E(z) \, dz, \tag{5}$$

where the dependence from the ion energy and fluence has been highlighted.

Since $I^E$ can be numerically calculated from a known profile $p^E(z)$ for the two ion energies employed in the implantations (values of 7.06 and 8.62 were found respectively for $E$ = 2 and 3 MeV), it is possible to fit the experimental *OPD* and *ALD* data with the real and the imaginary part of Eq.5, by employing the fluence $\phi$ as an independent variable and introducing a complex coefficient *c* as a fitting parameter, in a way that a same linear expression fits the ratio $COPD^E(\phi)/I^E$, irrespective of the ion energy.

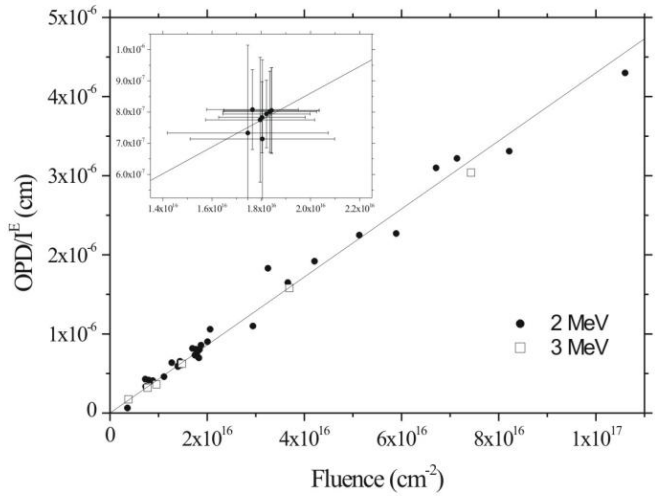

**Fig. 5**. Behaviour of the ratio $OPD^E(\phi)/I^E$ as a function of the fluency. In the inset, particular of the points representing eight different impantations at a same nominal fluency but with different values of the instantaneous current (a factor 5 of variation).

This is shown Figs. 5 and 6, where the ratio $COPD^E(\phi)/I^E$ is reported as a function of $\phi$, and is also confirmed by the data in Tab.I, where the coefficients Re(*c*) and Im(*c*), found by linear regression of the 2 and 3 MeV implantations data are shown to be compatible within the uncertainties.

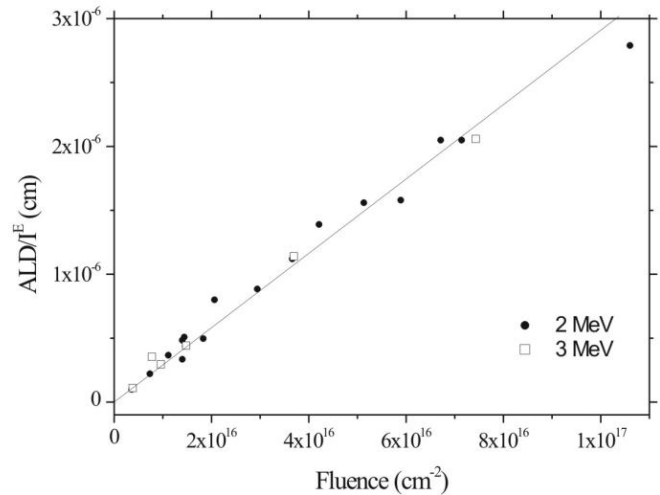

**Fig. 6**. Behaviour of the ratio $ALD^E(\phi)/I^E$ as a function of the fluence

Furthermore, the inset of Fig. 5 reports data relevant to eight implantations performed with 2 MeV protons at the same fluence with different ion currents (ranging from 0.2 nA to 1 nA). The OPD data are compatible within the experimental uncertainties, thus indicating that the fluence delivery rate has no significant effect on the refractive index variation.

|  | Re($c$) (cm$^3$) | Im($c$) (cm$^3$) |
|---|---|---|
| $E$ = 2 MeV | $(4.34\pm0.05)10^{-23}$ | $(2.86\pm0.05)10^{-23}$ |
| $E$ = 3 MeV | $(4.26\pm0.12)10^{-23}$ | $(2.85\pm0.10)10^{-23}$ |

**Table 1.** Values of the coefficient $c$ resulting from the linear fitting of the data relevant to 2 MeV and 3 MeV proton implantations

### *3.4 Multilayer model and validation of the phenomenological model*

As mentioned above, Eqs. 4 were derived under the assumption of the independence of the processes of refraction and absorption of the probe laser beam, and in particular that internal reflections due to the variation of the refractive index can be disregarded. To validate this hypothesis, we elaborated a model describing the propagation of the probe laser beam in diamond through a number of layers of different refractive indices and extinction coefficients, thus considering all processes of refraction and absorption associated with the variation of the complex refractive index in the implanted material by setting at the layers interfaces the appropriate boundary conditions of continuity of the electric field and of its derivative. We considered a simulation grid identical to that of the SRIM simulation, with a constant vacancy density $v_i = p^E(z_i)$ ($1 \leq i \leq 100$) and a complex refractive index $\hat{n}_i = \hat{n}_0 + c \cdot v_i$ for each layer, adopting for the complex parameter $c$ the values obtained in the previous section.

By comparing the amplitude and phase shift of the transmitted wave with the reference incident wave it is possible to estimate the values of the optical path difference and the absorption length difference, for each value of energy and fluence. The difference between the resulting estimations of $COPD^E(z)$ and those obtained from the previously described phenomenological never exceeds 1%, that is well bellow the experimental errors, confirming the validity of the approximation stated by eq. 4.

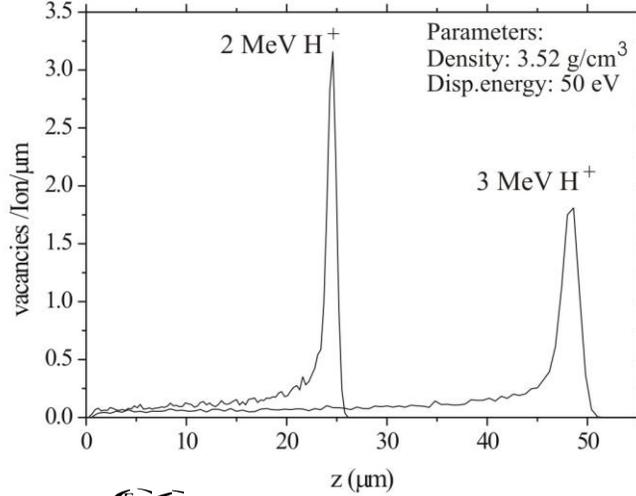

**Fig. 7.** Functions $p^E(z)$ for implantations of protons with energies of 2 and 3 MeV.

### 4. Discussion and conclusions

Interference microscopy and transmission spectroscopy has been exploited to study the dependence of diamond refractive index and extinction coefficient on the damage produced by 2-3 MeV H ions, for fluencies ranging over $10^{15}$-$10^{17}$ cm$^{-2}$. At these fluencies, the vacancy densities at end-of-range reaches $2.5 \cdot 10^{21}$ cm$^{-3}$, well below the amorphization threshold. Up to

these values, the dependence of the complex refractive index on the vacancy density ν results to be linear:

$$\hat{n} = 2.41 + [(4.34 \pm 0.05) + i \cdot (2.86 \pm 0.04)] \cdot 10^{-23} \text{cm}^3 \cdot v \quad (6)$$

This expression were deduced with the aid of a phenomenological model based on the integration along the probe beam path of the real and imaginary parts of the complex refractive index, under the assumptions that the interplay between the refraction and absorption processes can be neglected, as demonstrated *a posteriori* with a complete multi-layer model. The experimental results suggest that the variation of the refractive index depends only on the overall vacancy density induced by the radiation during the process, irrespectively of the ion energy and of the beam intensity.

Where a comparison is possible with previous reports about the optical characteristics of ion-damaged diamond, [18,20,22,23] the increasing trend of the real part of the refractive index is confirmed, and the linear coefficients, although determined with much higher uncertainty, are compatible with our results. In the very early report of ref. [18], the refractive index of diamond implanted with 20 keV $C^+$ ions exhibits a monotonic increase as a function of implantation fluence, with linear coefficients strongly dependent on the sample and ranging from about 2 to $10 \cdot 10^{-23}$ cm$^3$. The linear dependence holds up to a damage level at which the refractive index seems to saturate; such saturation level corresponds to a total atomic concentration of 0.025, i.e. $4.5 \cdot 10^{21}$ vacancies cm$^{-3}$, a value slightly exceeding the maximum damage density explored in the present work ($2.5 \cdot 10^{21}$ cm$^{-3}$). For one of the four diamond samples reported in ref.[18] (namely, sample I), the dependence of the refractive index from the damage density (estimated with the usual linear approximation from the damage profile of 20 keV C ions) is in very satisfactory agreement with our result, while other samples exhibited rather different trends. From such very early report it is not possible to reconstruct the types of the different diamond samples employed. Differently from what reported in ref.[46], in ref.[22] no clear trend emerges in the variation of the refractive index and therefore a direct comparison with the present work is difficult. In ref.[23] the authors report about a low value of the refractive index for the heavily damaged buried layers, whose damage-induced vacancy density amount to about $4 \cdot 10^{-22}$ cm$^{-3}$. In these conditions, the degree of amorphization/graphitization by far exceeds what reported in the present work. Finally, it is worth remarking that the results are in good agreement with more recent ellipsometric studies of the refractive index variation in shallow layers implanted with heavy ions at low damage densitities [20], for which consistent linearly increasing trends are reported.

The increasing trend of the refractive index as a function of induced damage is somewhat surprising with respect to what reported in other materials, such as quartz [46] or zircon [47], for example. This is because the most direct effect of ion implantation in crystals usually consists in the progressive amorphization of the substrate, which invariably leads to a decrease of the atomic density and therefore of the refractive index. Although often quantitatively predominant, the above-mentioned process is not the only effect determining a variation in refractive index. Beside volume expansion, other damage-related effects can occur which have a significant and direct effect on the refractive index, namely changes in atomic bond polarizability and structure factors, as expressed by the Wei adaptation of the Lorentz-Lorenz equation:

$$\frac{\Delta n}{n} = \frac{(n^2 - 1)(n^2 + 2)}{6n^2} \left[ -\frac{\Delta V}{V} + \frac{\Delta \alpha}{\alpha} + F \right] \quad (7)$$

where $V$ is volume, $\alpha$ is polarizability and $F$ is the structure factor of the target implanted material [48].

Although the volume expansion term is dominating in most cases, the structural modification results in changes of the chemical bonds and subsequently of the material polarizability. Such changes can be either positive or negative in sign and therefore, although the detailed analysis of these complex mechanisms goes beyond the scopes of the present work, it is reasonable to expect strong polarizability-related effects in a peculiar material such as diamond, in which the nature of the chemical bond can be subjected to drastic changes (i.e. from the strongly covalent $sp^3$ bonds to $sp^2$ bonds).

While for low damage levels (well below the amorphization threshold, as mentioned above), polarizability-related effects related to the formation of isolated $sp^2$ defects can dominate over the volume effects, it is reasonable to expect that at higher damage levels the amorphization of the diamond $sp^3$ lattice can lead to predominant density effects and thus to the reduction of the refractive index, as indeed observed in ref.[23].

We conclude by remarking that further investigation should be necessary to ascertain if the same mechanisms occur also for the damage induced by other ion species, but the present work indicates that a proton beam can be used in tailoring the optical properties of diamond in the MeV range with the help of a common damage simulation software such as SRIM. The methodology of measurements and analysis which we have adopted for this study is of ease and versatile use, for application for any transparent material within very large range of energies and fluences.

**Acknowledgments**

This work is supported by the "Accademia Nazionale dei Lincei—Compagnia di San Paolo" Nanotechnology grant and by "FARE" experiments of "Istituto Nazionale di Fisica Nucleare" (INFN), which are gratefully acknowledged.